\documentstyle{article}
\newcommand{\D}{\displaystyle}
\newcounter{bla}
\newenvironment{refnummer}{%
\list{[\arabic{bla}]}%
{\usecounter{bla}%
 \setlength{\itemindent}{0pt}%
 \setlength{\topsep}{0pt}%
 \setlength{\itemsep}{0pt}%
 \setlength{\labelsep}{2pt}%
 \setlength{\listparindent}{0pt}%
 \settowidth{\labelwidth}{[9]}%
 \setlength{\leftmargin}{\labelwidth}%
 \addtolength{\leftmargin}{\labelsep}%
 \setlength{\rightmargin}{0pt}}}
 {\endlist}

\makeatletter

\def\AmS{{\protect\the\textfont\tw@
  A\kern-.1667em\lower.5ex\hbox{M}\kern-.125emS}}
\makeatother

\begin{document}

\title{The Library of Subroutines for Calculation of Matrix Elements of Two--particle 
       Operators for Many--Electron Atoms\\}

\author{ Gediminas Gaigalas\\
        {\em Institute of Theoretical Physics and Astronomy,}\\
        {\em A. Go\v{s}tauto 12, Vilnius 2600, LITHUANIA}}

\maketitle

\begin{abstract}
In this paper a library for spin--angular integration in 
$LS$--coupling for many--electron atoms is presented.
The software is an implementation of a methodology based on the second
quantization in coupled tensorial form, the angular momentum theory in
3 spaces (orbital, spin and quasispin), and the graphical technique of
angular momentum. This implementation extends applications 
of the relevant methodology to open $f$-- shells
and leads to faster execution of angular integration codes.
The possibility of using some library routines for solving
various angular momentum problems in atomic physics is also discussed.
\end{abstract}

\vspace{6cm} {\em PACS:} 31.15.Ne, 31.25.-v, 32.10.-f, 32.30.-r

\vspace{0.3cm}

\newpage

{\bf PROGRAM SUMMARY} \\

\begin{small}

%\twocolumn

\noindent {\em Title of program} {\em :} SAI
{\em Catalogue identifier:} \\[10pt]
{\em Program obtainable from:} Institute of Theoretical Physics and Astronomy, 
A. Go\v{s}tauto 12, 2600 Vilnius, LITHUANIA.

E--mail: gaigalas@itpa.lt
  \\[10pt]
{\em Computer for which the library is designed and others on
which it has been tested:}\\
{\em Computers:} Pentium--based PCs; \\
{\em Installations:} Institute of Theoretical Physics and Astronomy, 
A. Go\v{s}tauto 12, 2600 Vilnius, LITHUANIA. \\[10pt]
{\em Programming language used in the new version:} FORTRAN 77~[1] \\[10pt]
%{\em Memory required to execute with typical data:} 64000K Bytes \hspace{18pt} words  \\[10pt]
{\em Peripherals used:} terminal, disk\\[10pt]
{\em No. of bits in a word:} 32\\[10pt]
{\em No. of processors used:} 1\\[10pt]
{\em Has the code been vectorised or parallelized?:} no \\[10pt]
{\em No. of bytes in distributed program, including test data, etc.:}
442 297 bytes \\[10pt]
{\em Distribution format:} compressed tar file \\[10pt]
%{\em CPC Program Library subprograms used:} none \\[10pt]
{\em Additional keywords :} atomic structure, configuration interaction,
reduced coefficients of fractional parentage, irreducible tensors,
angular momentum theory in three spaces (orbital, spin and quasispin),
second quantization in the coupled tensorial form, recoupling coefficients,
Slater integrals, complex atom, correlation,
wave functions, bound states, $LS$--coupling, $f$-- shell states.
\\[10pt]
{\em Nature of physical problem}\\
Theoretical determination of atomic orbitals, energy levels and radiative
transition data requires the calculation of
matrix elements of the relevant physical operators (see
the multiconfiguration Hartree--Fock method~[2], for example). 
The matrix elements of arbitrary operator can generally
be expressed as

$\lefteqn{\D
\sum_{i,j}~coef(i,j)~(\gamma_{i}L_{i}S_{i}||T^{(k_l k_s)}||\gamma_{j}L_{j}S_{j}),}$

where $T^{(k_l k_s)}$ is a tensorial operator of ranks $k_l$, $k_s$.
The program calculates the spin--angular part for
matrix elements $(\gamma_{i}L_{i}S_{i}||T^{(k_l k_s)}||\gamma_{j}L_{j}S_{j})$ of
one-- and/or two--particle operator $T^{(k_l k_s)}$.
\\[10pt]
{\em Method of solution} \\
This program is created involving the
angular methodology of [3--6]. It
has been extended to include partially filled $f$-- shells
in wave function expansions.
The classification of terms is  identical to that described in [6].
\\[10pt]
{\em Restrictions on the complexity of the problem}\\
The non--orthogonal orbitals are not supported.
\\[10pt]
%{\em Unusual features of the program }\\
%Some of the subroutines described in the libraries may be used separately,
%as electronic tables of standard quantities.
%\\[10pt]
{\em References}

%\begin{enumerate}
\begin{refnummer}
\item Microsoft, {\sl Fortran Powerstation Programmer's Guide} 
   (Microsoft Corporation, 1995).

\item C. Froese Fischer, T. Brage and P. J\" onsson,
   {\sl Computational Atomic Structure. An MCHF Approach}
   (Institute of Physics Publishing, Bristol/Philadelphia, 1997).

\item G. Gaigalas and Z. Rudzikas,
J. Phys. B: At. Mol. Phys. 29 (1996) 3303.

\item G. Gaigalas, Z. Rudzikas and C. Froese Fischer,
J. Phys. B: At. Mol. Phys. 30 (1997) 3747.

\item G. Gaigalas, A. Bernotas, Z. Rudzikas and C. Froese Fischer,
Physica Scripta 57 (1998) 207.

\item G. Gaigalas, Z. Rudzikas and C. Froese Fischer,
Atomic Data and Nuclear Data Tables 70 (1998) 1.

%\item C. Froese Fischer and B. Liu, Comput. Phys. Commun. 64 (1991) 406.
%\end{enumerate}
\end{refnummer}

\end{small}

%\onecolumn

\section*{LONG WRITE--UP}

\section{Introduction}

Models of many--electron atoms and ions require both relativistic and correlation
effects to be taken into account for getting very precise characteristics of atoms and
ions. This can be done, for example, by using multiconfiguration Haretree--Fock, 
configuration interaction method, various versions of perturbation 
theory, or semiempirical methods~\cite{R}. All
of them require the calculation of matrix elements of physical operators or 
effective operators from perturbation theory. The symmetry properties of atomic
states allow the calculation of matrix elements to be divided into the calculation 
of spin--angular terms and the accompanying radial integrals. The latter are the 
more straightforward and can be handled by methods such as used in the 
MCHF atomic structure package (\texttt{ATSP\_MCHF}) \cite{Fbook,F}. Such packages
have a modular structure, and the modules for calculation of the spin--angular part
for matrix elements of any operator can easily be replaced with new one, which is more
efficient for large scale computation of open shell atoms. 

\medskip

This paper describes modules of such a sort.
The new program (library for integration over spin--angular variables in atomic
theory -- library \texttt{SAI}) is based on combination of
the angular momentum theory, on the concept of irreducible tensorial sets, on a
generalized graphical approach, on the second quantization in coupled
tensorial form, on the quasispin approach and on standard quantities like the 
reduced coefficients of fractional parentage, completely reduced matrix elements of 
the unit tensors as well as a number of other completely reduced matrix elements 
occurring in various products of electron creation and annihilation operators~\cite{method6}. 
The program for calculation of standard quantities have been published as library
\texttt{SQ}~\cite{sai_sq}. The program presented in this paper
uses the library \texttt{SQ} for evaluation of these standard quantities.
The new module for the calculation of spin--angular coefficients is faster compared to
previous ones and can treat configurations with open $f^{N}$ shell. 

\medskip

The theoretical background is presented in Section 2, program organization 
is outlined in Section 3, the description of the use of
this library for other programs is given in Section 4. The program 
installation is presented in Section 5. Finally, a few examples are given in Section 6.

\section{Notations and Methodology of Angular Integrations}

\subsection{Matrix Elements Between Complex Configurations}

According to the approach \cite{method6,method2},
a general expression of a submatrix element
for any two--particle operator between functions with $u$ open shells, can be
written as follows:
\begin{eqnarray}
\label{eq:mg}
\lefteqn{\D
   (\psi _u\left( LS\right) ||G|| \psi _u\left( L^{\prime }S^{\prime}\right))}
   \nonumber  \\[1ex]
   & & =
   \sum_{n_il_i,n_jl_j,n_i^{\prime }l_i^{\prime },n_j^{\prime}l_j^{\prime }}
   (\psi _u\left( LS\right) ||\widehat{G}
   \left( n_il_i,n_jl_j,n_i^{\prime }l_i^{\prime },n_j^{\prime }l_j^{\prime } \right)
   ||\psi _u\left( L^{\prime }S^{\prime}\right) )
   \nonumber  \\[1ex]
   & & =
   \sum_{n_il_i,n_jl_j,n_i^{\prime }l_i^{\prime },n_j^{\prime}l_j^{\prime }}
   ~\sum_{\kappa _{12},\sigma _{12},\kappa_{12}^{\prime },\sigma _{12}^{\prime }}
   ~\sum_{K_l,K_s} \left( -1\right) ^\Delta \Theta^{\prime }\left( n_i\lambda _i,n_j\lambda 
   _j,n_i^{\prime }\lambda _i^{\prime},n_j^{\prime }\lambda _j^{\prime },\Xi \right) 
   \nonumber  \\[1ex]
   & & \times 
   T\left( n_i\lambda _i,n_j\lambda _j,n_i^{\prime }\lambda _i^{\prime },n_j^{\prime}
   \lambda _j^{\prime },\Lambda ^{bra},\Lambda ^{ket},\Xi ,\Gamma \right)
   ~R\left( \lambda _i,\lambda _j,\lambda _i^{\prime },\lambda _j^{\prime},\Lambda 
   ^{bra},\Lambda ^{ket},\Gamma \right),
\end{eqnarray}
where $\lambda \equiv l,s$, $\Lambda _l^{bra}\equiv
\left( L_i,L_j,L_i^{\prime },L_j^{\prime }\right) ^{bra}$, $\Lambda
_s^{bra}\equiv \left( S_i,S_j,S_i^{\prime },S_j^{\prime }\right) ^{bra}$
and $\Gamma$ refers to the array of coupling parameters connecting 
the recoupling matrix
$R\left( \lambda _i,\lambda _j,\lambda _i^{\prime },\lambda _j^{\prime },
\Lambda ^{bra},\Lambda ^{ket},\Gamma \right) $ to 
the submatrix element
$T\left( n_i\lambda _i,n_j\lambda _j,n_i^{\prime }\lambda _i^{\prime
},n_j^{\prime }\lambda _j^{\prime },\Lambda ^{bra},\Lambda ^{ket},\Xi
,\Gamma \right) $.
The expression (\ref{eq:mg})
has summation over intermediate ranks 
$\kappa _{12}$, $\sigma _{12}$, $\kappa_{12}^{\prime }$, $\sigma _{12}^{\prime }$,
$K_l$, $K_s$ in $T\left( n_i\lambda _i,n_j\lambda _j,n_i^{\prime }\lambda _i^{\prime
},n_j^{\prime }\lambda _j^{\prime },\Lambda ^{bra},\Lambda ^{ket},\Xi
,\Gamma \right) $.

\medskip

So, to calculate the spin--angular part of a submatrix element of
this type, one has to obtain:

\begin{enumerate}

\item Recoupling matrix
$R\left( \lambda _i,\lambda _j,\lambda _i^{\prime },\lambda _j^{\prime },
\Lambda ^{bra},\Lambda ^{ket},\Gamma \right) $. This recoupling matrix accounts
for the change in going from matrix element
$(\psi _u \left( LS\right) ||G|| \psi _u \left( L^{\prime }S^{\prime}\right))$,
which has $u$ open shells in the {\em bra} and {\em ket} functions, to the submatrix 
element
$T\left( n_i\lambda _i,n_j\lambda _j,n_i^{\prime }\lambda _i^{\prime
},n_j^{\prime }\lambda _j^{\prime },\Lambda ^{bra},\Lambda ^{ket},\Xi
,\Gamma \right) $, which has only the shells being acted upon by the
two--particle operator in its {\em bra} and {\em ket} functions.

\item Submatrix elements
$T\left( n_i\lambda _i,n_j\lambda _j,n_i^{\prime }\lambda _i^{\prime
},n_j^{\prime }\lambda _j^{\prime },\Lambda ^{bra},\Lambda ^{ket},\Xi
,\Gamma \right) $.

\item Phase factor $\Delta $.

\item $\Theta ^{\prime }\left(
n_i\lambda _i,n_j\lambda _j,n_i^{\prime }\lambda _i^{\prime },n_j^{\prime
}\lambda _j^{\prime },\Xi \right) $,
which is proportional to the radial part. It consists of
a submatrix element
$( n_i \lambda_i n_j \lambda_j ||g^{(\kappa_1 \kappa_2 k, \sigma_1 \sigma_2 k)}||
\psi _u \left( L^{\prime }S^{\prime}\right))$, and in some cases of simple factors and 
3$nj$--coefficients (for more details see \cite{method2}).

\end {enumerate}

\medskip

Some important points to note are the following:

\medskip

1. The recoupling matrices
$R\left( \lambda
_i,\lambda _j,\lambda _i^{\prime },\lambda _j^{\prime },\Lambda
^{bra},\Lambda ^{ket},\Gamma \right) $ in our approach are much simpler than
in other known approaches. We have obtained their analytical expressions in
terms of just $6j$-- and $9j$--coefficients. That is why we choose a special
form of operator in second quantization, where second quantization
operators
acting upon the same shell are tensorially coupled together.

\medskip

2.  The tensorial part of a two--particle operator is expressed in terms of
(products of) operators of the type
$A^{\left(
kk\right) }\left( n\lambda ,\Xi \right) $, $B^{\left( kk\right) }(n\lambda
,\Xi )$, $C^{\left( kk\right) }(n\lambda ,\Xi )$, $D^{\left( ls\right) }$, $%
E^{\left( kk\right) }(n\lambda ,\Xi )$.
Their explicit expressions are (\ref{eq:ta})--(\ref{eq:te}):

\begin{equation}
\label{eq:ta}
   a_{m_q}^{\left( q\lambda \right) },
\end{equation}

\begin{equation}
\label{eq:tb}
   \left[ a_{m_{q1}}^{\left( q\lambda \right) }\times
   a_{m_{q2}}^{\left( q\lambda \right) }\right] ^{\left( \kappa _1\sigma _1\right) },
\end{equation}

\begin{equation}
   \label{eq:tc}\left[ a_{m_{q1}}^{\left( q\lambda \right) }\times \left[
   a_{m_{q2}}^{\left( q\lambda \right) }\times a_{m_{q3}}^{\left( q \lambda \right)}
   \right] ^{\left( \kappa _1\sigma _1\right) }\right] ^{\left( \kappa _2\sigma _2\right) },
\end{equation}

\begin{equation}
\label{eq:td}
   \left[ \left[ a_{m_{q1}}^{\left( q\lambda \right) }\times
   a_{m_{q2}}^{\left( q\lambda \right) }\right] ^{\left( \kappa _1\sigma _1\right)}\times
   a_{m_{q3}}^{\left( q\lambda \right) }\right] ^{\left( \kappa _2\sigma _2\right) },
\end{equation}

\begin{equation}
\label{eq:te}
   \left[ \left[ a_{m_{q1}}^{\left( q\lambda \right) }\times
   a_{m_{q2}}^{\left( q\lambda \right) }\right] ^{\left( \kappa _1\sigma
   _1\right) }\times \left[ a_{m_{q3}}^{\left( q\lambda \right) }\times
   a_{m_{q4}}^{\left( q\lambda \right) }\right] ^{\left( \kappa _2\sigma
   _2\right) }\right] ^{\left( kk\right)} .
\end{equation}
We denote their submatrix elements by
$T\left( n_i\lambda _i,n_j\lambda _j,n_i^{\prime }\lambda _i^{\prime
},n_j^{\prime }\lambda _j^{\prime },\Lambda ^{bra},\Lambda ^{ket},\Xi
,\Gamma \right) $. The parameter $\Gamma $ represents
the whole array of parameters
connecting the recoupling matrix
$R\left( \lambda _i,\lambda _j,\lambda _i^{\prime },\lambda
_j^{\prime },\Lambda ^{bra},\Lambda ^{ket},\Gamma \right) $ to the submatrix
element
$T\left( n_i\lambda _i,n_j\lambda _j,n_i^{\prime }\lambda _i^{\prime
},n_j^{\prime }\lambda _j^{\prime },\Lambda ^{bra},\Lambda ^{ket},\Xi
,\Gamma \right) $. It is worth noting that each of the
tensorial quantities (\ref
{eq:ta})--(\ref{eq:te}) act upon one and the same shell. So, all the
advantages of tensor algebra and the quasispin formalism may be efficiently
exploited in the process of their calculation.

\medskip

We obtain the submatrix elements of operator (\ref{eq:ta}) by using
straightforwardly the Wigner--Eckart theorem in quasispin space:
\begin{eqnarray}
\label{eq:tf}
   \left( l^N\;\alpha QLS||a_{m_q}^{\left( qls\right) }||l^{N^{\prime}}\;\alpha 
   ^{\prime }Q^{\prime }L^{\prime }S^{\prime }\right) =-\left[Q\right] ^{-1/2}\left[
   \begin{array}{ccc}
   Q^{\prime } & 1/2 & Q \\
   M_Q^{\prime } & m_q & M_Q
   \end{array}
   \right] \left( l\;\alpha QLS|||a^{\left( qls\right) }|||l\;\alpha ^{\prime}Q^{\prime 
   }L^{\prime }S^{\prime }\right),
\end{eqnarray}
where the last multiplier in (\ref{eq:tf}) is the so--called
reduced coefficient of fractional parentage (RCFP)
and we use a shorthand notation $(2k+1)\cdot ...\equiv [k,...]$.
The details of the use of quasispin approach are discussed in monograph~\cite{R}.

\medskip

The value of a submatrix element of operator (\ref{eq:tb}) is obtained
by basing ourselves on (33), (34) in \cite{method1}.
In the other three cases
(\ref{eq:tc}), (\ref{eq:td}), (\ref{eq:te}) we obtain them by using (5.16) of
Rudzikas \cite{R}:
\begin{eqnarray}
\label{eq:tg}
\lefteqn{\D
   (nl^N\;\alpha QLS||\left[ F^{\left( \kappa _1\sigma _1\right) }\left(
   n\lambda \right) \times G^{(\kappa _2\sigma _2)}\left( n\lambda \right)
   \right] ^{\left( kk\right) }||nl^{N^{\prime }}\;\alpha ^{\prime }Q^{\prime
   }L^{\prime }S^{\prime })}
   \nonumber  \\[1ex]
   & & =
   \left( -1\right) ^{L+S+L^{\prime }+S^{\prime }+2k}\left[ k\right]
   \displaystyle {\sum_{\alpha ^{\prime \prime }Q^{\prime \prime }L^{\prime
   \prime }S^{\prime \prime }}}(nl^N\;\alpha QLS||F^{\left( \kappa _1\sigma
   _1\right) }\left( n\lambda \right) ||nl^{N^{\prime \prime }}\;\alpha
   ^{\prime \prime }Q^{\prime \prime }L^{\prime \prime }S^{\prime \prime })
   \nonumber  \\[1ex]
   & & \times
   (nl^{N^{\prime \prime }}\;\alpha ^{\prime \prime }Q^{\prime \prime }L^{\prime \prime 
   }S^{\prime \prime }||G^{(\kappa _2\sigma _2)}\left( n\lambda \right) ||nl^{N^{\prime 
   }}\;\alpha ^{\prime }Q^{\prime }L^{\prime }S^{\prime })
   \left\{
   \begin{array}{ccc}
   \kappa _1 & \kappa _2 & k \\
   L^{\prime } & L & L^{\prime \prime }
   \end{array}
   \right\} \left\{
   \begin{array}{ccc}
   \sigma _1 & \sigma _2 & k \\
   S^{\prime } & S & S^{\prime \prime }
   \end{array}
   \right\} ,
\end{eqnarray}
where $F^{\left( \kappa _1\sigma _1\right) }\left( n\lambda \right) $ and
$G^{(\kappa _2\sigma _2)}\left( n\lambda \right) $ is one of (\ref{eq:ta}) or
(\ref{eq:tb}) and the submatrix elements correspondingly are defined by (\ref
{eq:tf}) and (33), (34) in \cite{method1}. $N^{\prime \prime }$ is defined by
second quantization operators occurring in $F^{\left( \kappa _1\sigma
_1\right) }\left( n\lambda \right) $ and $G^{(\kappa _2\sigma _2)}\left(
n\lambda \right) $.

\medskip

As is seen, by using this approach, the calculation of the angular
parts of matrix elements between functions with $u$ open shells ends up as
a calculation of submatrix elements of tensors (\ref{eq:ta}), (\ref{eq:tb})
within single shell of equivalent electrons. As these completely reduced
submatrix elements
(reduced in the quasispin, orbital and spin spaces)
do not depend on the occupation number of the shell, the
tables for them are reduced considerably in comparison with the tables of
analogous submatrix elements of tensorial quantities $U^k,$ $V^{k_1k_2}$
(Jucys and Savukynas~\cite{JS} or Cowan~\cite{Cowan})
and the tables of fractional parentage
coefficients (CFP). That is why
the expressions obtained are very useful in practical calculations.
This is extremely important for the $f$-- shell, where the number of CFPs
for $f^1$ -- $f^{14}$ equals 54408~\cite{NK}, whereas the number of RCFPs,
taking into account the transposition symmetry
property of RCFP, is only 14161 -- of which only 3624 are nonzero~\cite{method5}.

\medskip

We do not present details on obtaining phase factors $\Delta $ and
$\Theta ^{\prime }\left( n_i\lambda _i,n_j\lambda _j,n_i^{\prime }\lambda
_i^{\prime },n_j^{\prime }\lambda _j^{\prime },\Xi \right) $, since no
essential generalizations may be made here; those are possible only after
a particular operator is chosen (for more details see 
\cite{method2,method3}).

\subsection{The Electrostatic Electron Interaction,
Spin--Spin and Spin--Other--Orbit Operators}

The {\it electrostatic (Coulomb) electron interaction} operator $H^{Coulomb}$
itself contains the tensorial structure
\begin{eqnarray}
\label{eq:co-aa}
\D
   H^{Coulomb} \equiv \sum_{k} H_{Coulomb}^{(kk0,000)}
\end{eqnarray}
and its submatrix element is:
\begin{eqnarray}
\label{eq:co-a}
   \left( n_i\lambda _in_j\lambda _j\left\| H_{Coulomb}^{(kk0,000)}\right\|
   n_{i^{\prime }}\lambda _{i^{\prime }}n_{j^{\prime }}\lambda _{j^{\prime }}\right)
   = 2 [k]^{1/2} \left( l_i\left\| C^{\left( k \right) }\right\| l_{i^{\prime }}\right)
   \left( l_j\left\| C^{\left( k \right) }\right\| l_{j^{\prime }}\right)
   R_{k}\left( n_il_i n_{i^{\prime }}l_{i^{\prime }},n_jl_j n_{j^{\prime
   }}l_{j^{\prime }}\right).
\end{eqnarray}

\medskip

The {\it spin--spin} operator $H^{ss}$ itself contains tensorial structure
of two different types, summed over $k$:
\begin{eqnarray}
\label{eq:ss-a}
   \D H^{ss} \equiv \sum_{k}
   \left[ H_{ss}^{(k+1 k-1 2,112)} + H_{ss}^{(k-1 k+1 2,112)} \right].
\end{eqnarray}
Their submatrix elements are:
\begin{eqnarray}
\label{eq:ss-b}
\lefteqn{
   \left( n_i\lambda _in_j\lambda _j\left\| H_{ss}^{\left( k+1k-12,112\right)
   }\right\| n_{i^{\prime }}\lambda _{i^{\prime }}n_{j^{\prime }}\lambda
   _{j^{\prime }}\right)}
   \nonumber  \\[1ex]
   & & =\frac 3{\sqrt{5}}\sqrt{\left( 2k+3\right) ^{\left( 5\right) }}
   \left( l_i\left\| C^{\left( k+1\right) }\right\| l_{i^{\prime }}\right)
   \left( l_j\left\| C^{\left( k-1\right) }\right\| l_{j^{\prime }}\right)
   N^{k-1}\left( n_il_in_jl_j,n_{i^{\prime }}l_{i^{\prime }}n_{j^{\prime
   }}l_{j^{\prime }}\right),
\end{eqnarray}
\begin{eqnarray}
\label{eq:ss-c}
\lefteqn{
   \left( n_i\lambda _in_j\lambda _j\left\| H_{ss}^{\left( k-1k+12,112\right)
   }\right\| n_{i^{\prime }}\lambda _{i^{\prime }}n_{j^{\prime }}\lambda
   _{j^{\prime }}\right)}
   \nonumber  \\[1ex]
   & & =\frac 3{\sqrt{5}}\sqrt{\left( 2k+3\right) ^{\left( 5\right) }}
   \left( l_i\left\| C^{\left( k-1\right) }\right\| l_{i^{\prime }}\right)
   \left( l_j\left\| C^{\left( k+1\right) }\right\| l_{j^{\prime }}\right)
   N^{k-1}\left( n_jl_jn_il_i,n_{j^{\prime }}l_{j^{\prime }}n_{i^{\prime
   }}l_{i^{\prime }}\right),
\end{eqnarray}
where we use a shorthand notation
$\left( 2k+3\right) ^{\left( 5\right) } \equiv \left( 2k+3\right)
\left( 2k+2\right)\left( 2k+1\right)\left( 2k\right)\left( 2k-1\right)$ and
radial integral (\ref{eq:ss-b}), (\ref{eq:ss-c}) is defined as in Glass and
Hibbert~\cite{GlassH}:
\begin{eqnarray}
\label{eq:ss-d}
   N^k\left( n_il_in_jl_j,n_{i^{\prime }}l_{i^{\prime }}n_{j^{\prime
   }}l_{j^{\prime }}\right) =\frac{\alpha ^2}4\int_0^\infty \int_0^\infty 
   P_i\left( r_1\right) P_j\left(
   r_2\right) \frac{r_2^k}{r_1^{k+3}}\epsilon (r_1-r_2)P_{i^{\prime }}\left(
   r_1\right) P_{j^{\prime }}\left( r_2\right) dr_1dr_{2},
\end{eqnarray}
where $\epsilon (x)$ is a Heaviside step--function,
\begin{eqnarray}
\label{eq:ss-e}
   \epsilon (x)=\left\{
   \begin{array}{ll}
   1 ; & \mbox{ for } x>0, \\ 0 ; & \mbox{ for } x\leq 0.
   \end{array}
   \right.
\end{eqnarray}

\medskip

The {\it spin--other--orbit} operator $H^{soo}$ itself contains tensorial
structure of six different types, summed over $k$:
\begin{eqnarray}
\label{eq:soo-a}
   \D H^{sso} \equiv \sum_{k}
   \left[ H_{sso}^{(k-1 k 1,101)} + H_{sso}^{(k-1 k 1,011)} + H_{sso}^{(k k 1,101)}
   + H_{sso}^{(k k 1,011)} + H_{sso}^{(k+1 k 1,101)} + H_{sso}^{(k+1 k 1,011)}
   \right].
\end{eqnarray}
Their submatrix elements are:
\begin{eqnarray}
\label{eq:soo-b}
\lefteqn{
   \left( n_i\lambda _in_j\lambda _j\left\| H_{soo}^{\left( k-1k1,\sigma
  _1\sigma _21\right) }\right\| n_{i^{\prime }}\lambda _{i^{\prime
   }}n_{j^{\prime }}\lambda _{j^{\prime }}\right)} 
   \nonumber  \\[1ex]
   & &  =
   2\cdot 2^{\sigma _2}\left\{ \left( 2k-1\right) \left( 2k+1\right) \right. 
   \left. \left( l_i+l_{i^{\prime }}-k+1\right) \left(
   k-l_i+l_{i^{\prime }}\right) \left( k+l_i-l_{i^{\prime }}\right) \left(
   k+l_i+l_{i^{\prime }}+1\right) \right\} ^{1/2} 
   \nonumber  \\[1ex]
   & & \times 
   \left( k\right) ^{-1/2}\left( l_i\left\| C^{\left( k\right) }\right\|
   l_{i^{\prime }}\right) \left( l_j\left\| C^{\left( k\right) }\right\|
   l_{j^{\prime }}\right) N^{k-2}\left( n_jl_jn_il_i,n_{j^{\prime
   }}l_{j^{\prime }}n_{i^{\prime }}l_{i^{\prime }}\right),
\end{eqnarray}
\begin{eqnarray}
\label{eq:soo-d}
\lefteqn{
   \left( n_i\lambda _in_j\lambda _j\left\| H_{soo}^{\left( k+1k1,\sigma
_1\sigma _21\right) }\right\| n_{i^{\prime }}\lambda _{i^{\prime
}}n_{j^{\prime }}\lambda _{j^{\prime }}\right)}
   \nonumber  \\[1ex]
   & &  =
   2\cdot 2^{\sigma _2}\left\{ \left( 2k+1\right) \left( 2k+3\right)
   \left( l_i+l_{i^{\prime }}-k\right) \left( k-l_i+l_{i^{\prime
   }}+1\right) \left( k+l_i-l_{i^{\prime }}+1\right) \left( k+l_i+l_{i^{\prime
   }}+2\right) \right\} ^{1/2}
   \nonumber  \\[1ex]
   & & \times
   \left( k+1\right) ^{-1/2}\left( l_i\left\| C^{\left( k\right)
   }\right\| l_{i^{\prime }}\right) \left( l_j\left\| C^{\left( k\right)
   }\right\| l_{j^{\prime }}\right) N^k\left( n_il_in_jl_j,n_{i^{\prime
   }}l_{i^{\prime }}n_{j^{\prime }}l_{j^{\prime }}\right) .
\end{eqnarray}
\begin{eqnarray}
\label{eq:soo-c}
\lefteqn{
   \left( n_i\lambda _in_j\lambda _j\left\| H_{soo}^{\left( kk1,\sigma _1\sigma
   _21\right) }\right\| n_{i^{\prime }}\lambda _{i^{\prime }}n_{j^{\prime
   }}\lambda _{j^{\prime }}\right)}
   \nonumber  \\[1ex]
   & &  =
   -2\cdot 2^{\sigma _2}\left( 2k+1\right)
   ^{1/2}\left( l_i\left\| C^{\left( k\right) }\right\| l_{i^{\prime }}\right)
   \left( l_j\left\| C^{\left( k\right) }\right\| l_{j^{\prime }}\right)
   \left\{ \left( k\left( k+1\right) \right) ^{-1/2} \right.
   \nonumber  \\[1ex]
   & & \times
   \left( l_i\left(
   l_i+1\right) -k\left( k+1\right) -l_{i^{\prime }}\left( l_{i^{\prime
   }}+1\right) \right) 
   \left\{ \left( k+1\right) N^{k-2}\left( n_jl_jn_il_i,n_{j^{\prime
   }}l_{j^{\prime }}n_{i^{\prime }}l_{i^{\prime }}\right) \right.
   \nonumber  \\[1ex]
   & & -
   \left. \left. kN^k\left(
   n_il_in_jl_j,n_{i^{\prime }}l_{i^{\prime }}n_{j^{\prime }}l_{j^{\prime
   }}\right) \right\} 
%   \nonumber  \\[1ex]
%
%   & &
   - 2\left( k\left( k+1\right) \right) ^{1/2}V^{k-1}\left(
   n_il_in_jl_j,n_{i^{\prime }}l_{i^{\prime }}n_{j^{\prime }}l_{j^{\prime
   }}\right) \right\}.
\end{eqnarray}
The radial integrals in  (\ref{eq:soo-b})--(\ref{eq:soo-c}) are defined by
(\ref{eq:ss-d}) and below (see
Glass and Hibbert~\cite{GlassH}):
\begin{eqnarray}
\label{eq:m-k}
   V^k\left( n_il_in_jl_j,n_{i^{\prime }}l_{i^{\prime }}n_{j^{\prime
   }}l_{j^{\prime }}\right) = \frac{\alpha ^2}4\int_0^\infty \int_0^\infty
   P_i\left( r_1\right) P_j\left( r_2\right) \frac{r_{<}^{k-1}}{r_{>}^{k+2}}r_2
   \frac \partial {\partial r_1}P_{i^{\prime }}\left( r_1\right) 
   P_{j^{\prime }}\left( r_2\right) dr_1dr_2.
\end{eqnarray}

\medskip

Now we have all we need (the operators with tensorial structure and their
submatrix elements) for obtaining the values of a matrix element of these
operators for any number of open shells in bra and ket functions.
This lets us exploit all advantages of the approach by \cite{method2}.

\medskip

The spin--spin and spin--other--orbit operators themselves generally contain
tensorial structure of several different types. Therefore the expression
(\ref{eq:mg}) must be used separately for each possible tensorial structure
for performing spin--angular integrations according to \cite{method2}.
Each type of tensorial structure is associated with a different type
of recoupling matrix
$R\left( \lambda _i,\lambda _j,\lambda _i^{\prime },\lambda _j^{\prime },
\Lambda ^{bra},\Lambda ^{ket},\Gamma \right) $ and with different matrix
elements of standard tensorial quantities
$T\left( n_i\lambda _i,n_j\lambda _j,n_i^{\prime }\lambda _i^{\prime
},n_j^{\prime }\lambda _j^{\prime },\Lambda ^{bra},\Lambda ^{ket},\Xi
,\Gamma \right) $.

\medskip

The one--particle operators are treated in a similar manner. Their
expressions are much simpler and therefore we do not present them here, for
brevity. They may be found in \cite{gfg}.

\section{Description of the Library}

The library \texttt{SAI} presented in this paper is
aimed at the spin--angular integration for any one-- and 
two--particle operator.
It is a separate unit and
can easily be adapted to existing codes such as 
the MCHF atomic structure package (\texttt{ATSP\_MCHF}) \cite{Fbook,F}
or can easily be used to create a new one.
It contains four modules --
\texttt{SAI\_RECLS}, \texttt{SAI\_SQ},
\texttt{SAI\_NORE} and \texttt{SAI\_DUDU}.
They are classified according to the
methodology presented in papers
\cite{method6,method2,method1,method3},
and adhere to the principles of modular programming
(although FORTRAN 77 does not fully support this).

\medskip

The module \texttt{SAI\_SQ} was published separately~\cite{sai_sq}.
In this paper we will discuss briefly the new modules and subroutines contained therein.
The author does not attempt to describe in detail all the subroutines
belonging to separate libraries. In order to give the reader a more complete
view of the implementation of methodology, published in papers
\cite{method6, method2, method1, method5, method3, method7}, 
%\cite{method1} -- \cite{method7},
only the main
subroutines are described. Special attention is given to the description of
subroutines that other programs must call for 
use of this library.
The subroutines described in more detail
can easily be used separately from the complete library \texttt{SAI}, too.
These subroutines are useful for
creating new programs or modifying old ones, even
those based on the traditional methodology of angular calculations described
by Fano \cite{Fano}. 

\subsection{SAI\_RECLS}

This library contains 20 routines for calculation of recoupling matrices
\begin{eqnarray}
\label{eq:rec-a}
   R\left( \lambda _i,\lambda _j,\lambda _i^{\prime },\lambda _j^{\prime },\Lambda
   ^{bra},\Lambda ^{ket},\Gamma \right) = R\left( l_i,l_j,l_i^{\prime },l_j^{\prime
   },\Lambda _l^{bra},\Lambda _l^{ket},\Gamma _l\right) R\left(
   s,s,s,s,\Lambda _s^{bra},\Lambda _s^{ket},\Gamma _s\right).
\end{eqnarray}

\medskip

For more details see \cite{method2} (Section 4).
Most of the subroutines from this module use common blocks \texttt{CONSTS} 
and \texttt{MEDEFN} from
\texttt{ATSP\_MCHF}~\cite{Fbook,F}.

%\begin{tabbing}
\medskip

{\bf DLSA}n, n=1, 2, ... 6, evaluates respectively coefficients $C_{1}$, $C_{5}$,
$C_{2}$, $C_{4}$, $C_{3}$, $C^{\prime}_{6}$, defined in equations (15), (16), (23),
(21), (17), (25) of \cite{method2}.

\medskip

{\bf RECOUP0} checks the angular momentum selection rules
for the recoupling coefficients. For example, it uses the
expression (18) in one interacting shell case (see \cite{method2}).\\

\medskip

{\bf RECOUP}n, n=2, 3, 4, evaluates the Kronecker delta functions or calculates the recoupling 
coefficients respectively for the scalar operator
\begin{eqnarray}
\label{eq:rec-b}
   \left[ A^{(k)} \left( n_{1}l_{1} \right) \times B^{(k)} \left(
   n_{2}l_{2} \right) \right] ^{(0)},
\end{eqnarray}
\begin{eqnarray}
\label{eq:rec-c}
   \left[ \left[ A^{(k_1)} \left( n_{i}l_{i} \right) \times B^{(k_2)} \left(
   n_{j}l_{j} \right) \right] ^{(k)} \times C^{(k)} \left( n_{m}l_{m} \right)
   \right] ^{(0)},
\end{eqnarray}
\begin{eqnarray}
\label{eq:rec-d}
   \left[ \left[ A^{(k_1)} \left( n_{1}l_{1} \right) \times B^{(k_2)} \left(
   n_{2}l_{2} \right) \right] ^{(k)} \times
   \left[ C^{(k_3)} \left( n_{3}l_{3} \right) \times D^{(k_4)} \left( n_{4}l_{4}
   \right) \right] ^{(k)} \right] ^{(0)},
\end{eqnarray}
defined in equations (22), (26), (33) of \cite{method2}.
The $A^{(k)}$, $B^{(k_2)}$, $C^{(k_3)}$
and $D^{(k_4)}$ may be simple or composite tensorial operators.

\medskip

{\bf RLSP0} evaluates the Kronecker delta functions
$\delta \left ( L_i, L^{\prime}_i \right)$
for one and two interacting shells
(see (14) and (19) in \cite{method2}).\\

\medskip

{\bf RLSP00} evaluates the Kronecker delta functions
$\delta \left ( L_{i},L^{\prime}_i \right)$
for three and four interacting shells
(see (24) and (27) in \cite{method2}).\\

\medskip

{\bf RLSP}n, n=1, 2, 3 evaluates the Kronecker delta functions or calculates the recoupling 
coefficients respectively for the non--scalar operator
\begin{eqnarray}
\label{eq:rec-pa}
A^{(k)} \left( n_{1}l_{1} \right),
\end{eqnarray}
\begin{eqnarray}
\label{eq:rec-pb}
\left[ A^{(k_1)} \left( n_{1}l_{1} \right) \times B^{(k_2)} \left(
n_{2}l_{2} \right) \right] ^{(k)},
\end{eqnarray}
\begin{eqnarray}
\label{eq:rec-pc}
\left[ \left[ A^{(k_1)} \left( n_{i}l_{i} \right) \times B^{(k_2)} \left(
n_{j}l_{j} \right) \right] ^{(k_3)} \times C^{(k_4)} \left( n_{m}l_{m} \right)
\right] ^{(k)},
\end{eqnarray}
defined in equations (14), (19), (24) of \cite{method2}.
This routine
evaluates the Kronecker delta functions or calculates respectively the first or second part of 
recoupling coefficients for the non--scalar operator
\begin{eqnarray}
\label{eq:rec-pd}
\left[ \left[ A^{(k_1)} \left( n_{1}l_{1} \right) \times B^{(k_2)} \left(
n_{2}l_{2} \right) \right] ^{(k_3)} \times
\left[ C^{(k_4)} \left( n_{3}l_{3} \right) \times D^{(k_5)} \left( n_{4}l_{4}
\right) \right] ^{(k_3)} \right] ^{(k)}.
\end{eqnarray}
in the case n=4a, 4b.

\subsection{SAI\_NORE}

This library is for calculating the spin--angular parts of matrix elements for a
scalar two--particle operator. It contains 18 subroutines. 
Most of the subroutines from this module use common blocks \texttt{CONSTS} 
and \texttt{MEDEFN} from \texttt{ATSP\_MCHF} \cite{Fbook,F}.
%The calculations are
%performed according to the methodology \cite{method2}.

\medskip

From (\ref{eq:mg}) we see that the matrix element of any two--particle operator
can be written as a sum over all possible sets of active shell quantum numbers
$n_{i} l_{i}$. The systematic analysis of \cite{method2} aims to minimize the 
number of distributions, which is necessary to obtain the matrix elements of any
two--electron operator, when the bra and ket functions consist of arbitrary
number of shells.
Table \ref{israis} lists all these distributions and at the same time 
the 
expressions used by each of the subroutines
{\bf NONRELAT1}, {\bf NONRELAT2}, {\bf NONRELAT31}, {\bf NONRELAT32}, 
{\bf NONRELAT33}, {\bf NONRELAT41},
{\bf NONRELAT51}, {\bf NONRELAT52}, {\bf NONRELAT53}. The numbering of expressions is the
same as in paper \cite{method2}, where all these expressions are presented.
As the structure of all the subroutines mentioned earlier is the same, and
only different expressions are used and different subroutines are called,
we will discuss in more detail only one of these subroutines.

\begin{table}
\begin{center}
\caption{Scheme of the expressions for matrix elements of two--particle
scalar operator (like Coulomb interaction).}
\label{israis}
\begin{tabular}{ r c c c c c c c c c } \hline
{\bf Dis.} & {\bf $\widehat{G}$ } & {\bf $\widehat{G}(T)$ }
& {\bf $\alpha$ } & {\bf $\beta$ }
& {\bf $\gamma$ } & {\bf $\delta$ }& $ \tilde \Theta$ &
{\bf $R$ }& $\Delta$ \\ \hline \hline
\multicolumn{3}{l}{\sl NONRELAT1 }
       &    &   &  &  &  & & \\
$\alpha\alpha\alpha\alpha$ &(47)&(5)&(38),(35)& --& --& --&(48),(49)&(18)&(41)\\
$\alpha\beta\alpha\beta$   &(50)&(6)&(35)&(35)& --& --&(51)&(22)&(41)\\
$\beta\alpha\beta\alpha$   &(50)&(6)&(35)&(35)& --& --&(51)&(22)&(41)\\
$\alpha\beta\beta\alpha$   &(54)&(6)&(35)&(35)& --& --&(55)&(22)&(41)\\
$\beta\alpha\alpha\beta$   &(54)&(6)&(35)&(35)& --& --&(55)&(22)&(41)\\
\multicolumn{3}{l}{\sl NONRELAT2 }
       &    &   &  &  &  & & \\
$\alpha\alpha\beta\beta$   &(52)&(6)&(35)&(35)& --& --&(53)&(22)&(41)\\
\multicolumn{3}{l}{\sl NONRELAT31 }
       &    &   &  &  &  & & \\
$\beta\alpha\alpha\alpha$   &(56)&(6)&(36)&(34)& --& --&(58)&(22)&(42)\\
$\alpha\beta\alpha\alpha$   &(56)&(6)&(36)&(34)& --& --&(59)&(22)&(42)\\
\multicolumn{3}{l}{\sl NONRELAT32 }
       &    &   &  &  &  & & \\
$\beta\beta\beta\alpha$   &(60)&(6)&(34)&(37)& --& --&(62)&(22)&(42)\\
$\beta\beta\alpha\beta$   &(60)&(6)&(34)&(37)& --& --&(63)&(22)&(42)\\
\multicolumn{3}{l}{\sl NONRELAT33 }
       &    &   &  &  &  & & \\
$\beta\gamma\alpha\gamma$   &(50)&(7)&(34)&(34)&(35)& --&(51)&(26)&(42)\\
$\gamma\beta\gamma\alpha$   &(50)&(7)&(34)&(34)&(35)& --&(51)&(26)&(42)\\
$\gamma\beta\alpha\gamma$   &(54)&(7)&(34)&(34)&(35)& --&(55)&(26)&(42)\\
$\beta\gamma\gamma\alpha$   &(54)&(7)&(34)&(34)&(35)& --&(55)&(26)&(42)\\
\multicolumn{3}{l}{\sl NONRELAT41 }
       &    &   &  &  &  & & \\
$\gamma\gamma\alpha\beta$   &(52)&(7)&(34)&(34)&(35)& --&(53)&(26)&(42)\\
$\gamma\gamma\beta\alpha$   &(52)&(7)&(34)&(34)&(35)& --&(53)&(26)&(42)\\
$\alpha\beta\gamma\gamma$   &(52)&(7)&(34)&(34)&(35)& --&(53)&(26)&(42)\\
$\beta\alpha\gamma\gamma$   &(52)&(7)&(34)&(34)&(35)& --&(53)&(26)&(42)\\
\multicolumn{3}{l}{\sl NONRELAT51 }
       &    &   &  &  &  & & \\
$\alpha\beta\gamma\delta$   &(52)&(8)&(34)&(34)&(34)&(34)&(53)&(33)&(43)\\
$\beta\alpha\gamma\delta$   &(52)&(8)&(34)&(34)&(34)&(34)&(53)&(33)&(43)\\
$\alpha\beta\delta\gamma$   &(52)&(8)&(34)&(34)&(34)&(34)&(53)&(33)&(43)\\
$\beta\alpha\delta\gamma$   &(52)&(8)&(34)&(34)&(34)&(34)&(53)&(33)&(43)\\
$\gamma\delta\alpha\beta$   &(52)&(8)&(34)&(34)&(34)&(34)&(53)&(33)&(43)\\
$\gamma\delta\beta\alpha$   &(52)&(8)&(34)&(34)&(34)&(34)&(53)&(33)&(43)\\
$\delta\gamma\alpha\beta$   &(52)&(8)&(34)&(34)&(34)&(34)&(53)&(33)&(43)\\
$\delta\gamma\beta\alpha$   &(52)&(8)&(34)&(34)&(34)&(34)&(53)&(33)&(43)\\
% \hline
%\end{tabular}
%\end{center}
%\end{table}
%\begin{table}
%\vspace {0.5in} Table 3 (continued) \vspace {0.1in}
%\begin{center}
%\caption{Scheme of the expressions for matrix elements of any two--particle
%operator}
%\begin{tabular}{ r c c c c c c c c c } \hline
%{\bf Dis.} & {\bf $\widehat{G}$ } & {\bf $\widehat{G}(T)$ }
%& {\bf $\alpha$ } & {\bf $\beta$ }
%& {\bf $\gamma$ } & {\bf $\delta$ }& $ \tilde \Theta$ &
%{\bf $R$ }& $\Delta$ \\ \hline \hline
\multicolumn{3}{l}{\sl NONRELAT52 }
       &    &   &  &  &  & & \\
$\alpha\gamma\beta\delta$   &(50)&(8)&(34)&(34)&(34)&(34)&(51)&(33)&(43)\\
$\alpha\gamma\delta\beta$   &(54)&(8)&(34)&(34)&(34)&(34)&(55)&(33)&(43)\\
$\gamma\alpha\delta\beta$   &(50)&(8)&(34)&(34)&(34)&(34)&(51)&(33)&(43)\\
$\gamma\alpha\beta\delta$   &(54)&(8)&(34)&(34)&(34)&(34)&(55)&(33)&(43)\\
$\beta\delta\alpha\gamma$   &(50)&(8)&(34)&(34)&(34)&(34)&(51)&(33)&(43)\\
$\delta\beta\gamma\alpha$   &(50)&(8)&(34)&(34)&(34)&(34)&(51)&(33)&(43)\\
$\beta\delta\gamma\alpha$   &(54)&(8)&(34)&(34)&(34)&(34)&(55)&(33)&(43)\\
$\delta\beta\alpha\gamma$   &(54)&(8)&(34)&(34)&(34)&(34)&(55)&(33)&(43)\\
\multicolumn{3}{l}{\sl NONRELAT53 }
       &    &   &  &  &  & & \\
$\alpha\delta\beta\gamma$   &(50)&(8)&(34)&(34)&(34)&(34)&(51)&(33)&(43)\\
$\delta\alpha\gamma\beta$   &(50)&(8)&(34)&(34)&(34)&(34)&(51)&(33)&(43)\\
$\alpha\delta\gamma\beta$   &(54)&(8)&(34)&(34)&(34)&(34)&(55)&(33)&(43)\\
$\delta\alpha\beta\gamma$   &(54)&(8)&(34)&(34)&(34)&(34)&(55)&(33)&(43)\\
$\beta\gamma\alpha\delta$   &(50)&(8)&(34)&(34)&(34)&(34)&(51)&(33)&(43)\\
$\gamma\beta\delta\alpha$   &(50)&(8)&(34)&(34)&(34)&(34)&(51)&(33)&(43)\\
$\beta\gamma\delta\alpha$   &(54)&(8)&(34)&(34)&(34)&(34)&(55)&(33)&(43)\\
$\gamma\beta\alpha\delta$   &(54)&(8)&(34)&(34)&(34)&(34)&(55)&(33)&(43)\\
\hline
\end{tabular}
\end{center}
\end{table}

{\bf NONRELAT1} is meant for finding spin--angular coefficients for the distributions
$\alpha \alpha \alpha \alpha$, $\alpha \beta \alpha \beta$ and
$\beta \alpha \beta \alpha$.

In the $\alpha \alpha \alpha \alpha$ case, the program uses expression
(5) from \cite{method2}. In this case
\begin{eqnarray}
\label{eq:no-a}
\D
   \widehat{G}\left( I\right) \sim  \sum_{\kappa _{12},\sigma _{12},\kappa 
   _{12}^{\prime },\sigma _{12}^{\prime }}\sum_p \left[
   \Theta_{IIa} \left( n\lambda ,\Xi \right)
   A_{a}^{\left( 00\right) }\left( n\lambda ,\Xi \right) +
   \Theta_{IIb} \left( n\lambda ,\Xi \right)
   A_{(b)p,-p}^{\left( kk\right) }\left( n\lambda ,\Xi \right) \right],
\end{eqnarray}
\noindent where
\begin{eqnarray}
\label{eq:no-b}
   A_{a}^{\left( 00\right) }\left( n\lambda ,\Xi \right)=
   \left[ \left[ a^{\left( l_\alpha s\right) }\times \tilde a^{\left(
   l_\alpha s\right) }\right] ^{\left( \kappa _1\sigma _1\right) }\times \left[
   \tilde a^{\left( l_\alpha s\right) }\times a^{\left( l_\alpha s\right)
   }\right] ^{\left( \kappa _2\sigma _2\right) }\right] ^{\left(00\right)},
\end{eqnarray}
\begin{eqnarray}
\label{eq:no-c}
A_{(b)p,-p}^{\left( kk\right) }\left( n\lambda ,\Xi \right)=
\left[ a^{\left( l_\alpha s\right) }\times \tilde a^{\left( l_\alpha
s\right) }\right] _{p,-p}^{\left( kk \right) },
\end{eqnarray}
\noindent and
\begin{eqnarray}
\label{eq:no-d}
\lefteqn{
   \Theta_{IIa} \left( n\lambda ,\Xi \right) =
   \tilde \Theta _{IIa}\left( n_\alpha \lambda _\alpha ,n_\alpha \lambda
   _\alpha ,n_\alpha \lambda _\alpha ,n_\alpha \lambda _\alpha ,\Xi \right)}
   \nonumber  \\[1ex]
   & & =
   \frac 12\left( -1\right) ^{k-p}\left[ \kappa _1,\sigma _1,\kappa _2,\sigma
   _2\right] ^{-1/2} \left( n_\alpha \lambda _\alpha n_\alpha \lambda _\alpha ||g^{\left(
   \kappa _1\kappa _2k,\sigma _1\sigma _2k\right) }||n_\alpha \lambda _\alpha 
   n_\alpha \lambda _\alpha \right)
\end{eqnarray}
\noindent and
\begin{eqnarray}
\label{eq:no-e}
\lefteqn{
   \Theta_{IIb} \left( n\lambda ,\Xi \right)=
   \tilde \Theta _{IIb}\left( n_\alpha \lambda _\alpha ,n_\alpha \lambda
   _\alpha ,n_\alpha \lambda _\alpha ,n_\alpha \lambda _\alpha ,\Xi \right)}
   \nonumber  \\[1ex]
   & & =
   \left( -1\right) ^{k-p+1}\left( n_\alpha \lambda _\alpha n_\alpha \lambda
   _\alpha ||g^{\left( \kappa _1\kappa _2k,\sigma _1\sigma _2k\right)
   }||n_\alpha \lambda _\alpha n_\alpha \lambda _\alpha \right) \left\{
   \begin{array}{ccc}
   \kappa _1 & \kappa _2 & k \\
   l_\alpha & l_\alpha & l_\alpha
   \end{array}
   \right\} \left\{
   \begin{array}{ccc}
   \sigma _1 & \sigma _2 & k \\
   s & s & s
   \end{array}
   \right\} .
\end{eqnarray}

\medskip

The value of the reduced matrix element of operator (\ref{eq:no-b}) is
found by subroutine {\bf WWLS1}, and that of
(\ref{eq:no-c}) by {\bf W1} (see module \texttt{SAI\_SQ} in~\cite{sai_sq}).
The value of coefficient $\Theta_{IIa} \left( n\lambda ,\Xi \right)$
is calculated by {\bf COULOMBLS} subroutine, because this coefficient, to the
accuracy of a factor and a phase, is equal to that part of the
two--electron submatrix elements of electrostatic interaction,
which this subroutine is calculating. The coefficient
$\Theta_{IIb} \left( n\lambda ,\Xi \right)$ is found by {\bf COULOMBLS} and 
{\bf SIXJ},
and the recoupling matrix is investigated by
{\bf RECOUP0} (see section {\bf SAI\_RECLS}).

\medskip

For the distributions $\alpha \beta \alpha \beta$ and
$\beta \alpha \beta \alpha$
the subroutine {\bf NONRELAT1} uses
(6) of \cite{method2}, keeping in mind that
$\Theta \left( n_\alpha \lambda _\alpha ,n_\beta \lambda _\beta ,\Xi \right)$
is expressed as (51) of \cite{method2} and tensorial parts
$B^{\left( \kappa _{12}\sigma _{12}\right) }\left( n_\alpha \lambda _\alpha
,\Xi \right)$,
$C^{\left( \kappa _{12}^{\prime }\sigma _{12}^{\prime
}\right) }\left( n_\beta \lambda _\beta ,\Xi \right)$ are equal to (35) from
\cite{method2}. The coefficients
$\Theta \left( n_\alpha \lambda _\alpha ,n_\beta \lambda _\beta ,\Xi \right)$
are investigated by {\bf COULOMBLS}, the coefficients
$B^{\left( \kappa _{12}\sigma _{12}\right) }\left( n_\alpha \lambda _\alpha
,\Xi \right)$ and
$C^{\left( \kappa _{12}^{\prime }\sigma _{12}^{\prime
}\right) }\left( n_\beta \lambda _\beta ,\Xi \right)$ are found by {\bf W1W2LS}
from the \texttt{SAI\_SQ} library~\cite{sai_sq}, and the recoupling matrix is
calculated by {\bf RECOUP2}.

\medskip

For the distributions $\alpha \beta \beta \alpha$ and
$\beta \alpha \alpha \beta$
the subroutine {\bf NONRELAT1} uses
(6) of \cite{method2}, keeping in mind that
$\Theta \left( n_\alpha \lambda _\alpha ,n_\beta \lambda _\beta ,\Xi \right)$
is expressed as (55) of \cite{method2} and tensorial parts
$B^{\left( \kappa _{12}\sigma _{12}\right) }\left( n_\alpha \lambda _\alpha
,\Xi \right)$,
$C^{\left( \kappa _{12}^{\prime }\sigma _{12}^{\prime
}\right) }\left( n_\beta \lambda _\beta ,\Xi \right)$ are equal to (35) from
\cite{method2}. The coefficients
$\Theta \left( n_\alpha \lambda _\alpha ,n_\beta \lambda _\beta ,\Xi \right)$
are investigated by {\bf COULOMBLS} and {\bf SIXJ}, the coefficients
$B^{\left( \kappa _{12}\sigma _{12}\right) }\left( n_\alpha \lambda _\alpha
,\Xi \right)$ and
$C^{\left( \kappa _{12}^{\prime }\sigma _{12}^{\prime
}\right) }\left( n_\beta \lambda _\beta ,\Xi \right)$ are found by {\bf W1W2LS}
from the \texttt{SAI\_SQ} library~\cite{sai_sq}, and the recoupling matrix is
calculated by {\bf RECOUP2}.

\medskip

{\bf COULOMBLS} Investigates
the two--electron submatrix elements of electrostatic interaction
\begin{eqnarray}
   \left( n_i\lambda _in_j\lambda _j||g_{Coulomb}^{\left( kk0,000\right)
   }||n_i^{\prime }\lambda _i^{\prime }n_j^{\prime }\lambda _j^{\prime }\right),
\end{eqnarray}
according to the formula (9) of \cite{method1}. The values of these matrix
elements are needed because of (see  \cite{method2})
\begin{eqnarray}
\label{eq:no-aa}
   \Theta \left( \Xi \right) \sim \left( n_i\lambda _in_j\lambda
   _j\left\| g\right\| n_{i^{\prime }}\lambda _{i^{\prime }}n_{j^{\prime
   }}\lambda _{j^{\prime }}\right) .
\end{eqnarray}

\medskip

The value of the output parameter AA of this subroutine is:
\begin{eqnarray}
   AA=2\left[ k\right] ^{1/2}\left( l_i||C^{\left( k\right) }||l_i^{\prime
   }\right) \left( l_j||C^{\left( k\right) }||l_j^{\prime }\right).
\end{eqnarray}
%\begin{center}
%*   *  *
%\end{center}

\subsection{SAI\_DUDU}

This library is meant for the calculation of matrix elements of any 
one-- or two--particle operator. It contains 44 subroutines. 
Most of the subroutines from this module use common blocks \texttt{CONSTS} 
and \texttt{MEDEFN} from package \texttt{ATSP\_MCHF}~\cite{Fbook,F}.
Similar to the
\texttt{SAI\_NORE} library, it uses the methodology described in paper
\cite{method2}. Therefore the arrangement of library \texttt{SAI\_DUDU}
is analogous to that of library \texttt{SAI\_NORE}. Therefore we will not
go into computational details of spin--angular parts of one-- or two--particle
operator matrix elements, but instead will reserve all the attention to a
demonstration of connection of the Breit--Pauli operators to the general
algorithm of one-- or two--particle operator calculation.

\subsubsection{Spin--own--orbit interaction}

The subroutine {\bf SPINOR} investigates the
submatrix elements of the spin--own--orbit interaction operator
\begin{eqnarray}
   \left( n_i\lambda _i ||f_{s-o}^{\left( 11 \right) }||
   n_j^{\prime }\lambda _j^{\prime }\right),
\end{eqnarray}
according to the formula (5) of \cite{method5}.

\subsubsection{Spin--spin interaction}

The subroutine {\bf SSC} investigates the spin--spin operator, which
has the tensorial form (\ref{eq:ss-a})
%\begin{eqnarray}
%   H^{ss} \equiv \displaystyle {\sum_{ k }}
%   \left\{ H_{ss}^{(k+1 k-1 2, 1 1 2)} + H_{ss}^{(k-1 k+1 2, 1 1 2)} \right\},
%\end{eqnarray}
and finds the submatrix element of this operator between functions
with any number of open shells.
This subroutine is used for all distributions except
$\alpha\alpha\alpha\alpha$, $\alpha\beta\alpha\beta$,
$\beta\alpha\beta\alpha$, $\alpha\beta\beta\alpha$,
$\beta\alpha\alpha\beta$ and $\alpha\alpha\beta\beta$.
In the latter cases, instead of subroutine {\bf SSC} the subroutines
{\bf SS1111}, {\bf SS1212}, {\bf SS1212}, {\bf SS1221},
{\bf SS1221} and {\bf SS1122} are used.

\medskip

The subroutine {\bf SSA} investigates the
submatrix elements of the spin--spin interaction operator
\begin{eqnarray}
   \left( n_i\lambda _in_j\lambda _j||H_{ss}^{\left( k+1 k-12,112 \right)
   }||n_i^{\prime }\lambda _i^{\prime }n_j^{\prime }\lambda _j^{\prime }\right)
\end{eqnarray}
and
\begin{eqnarray}
   \left( n_i\lambda _in_j\lambda _j||H_{ss}^{\left( k-1 k+12,112 \right)
   }||n_i^{\prime }\lambda _i^{\prime }n_j^{\prime }\lambda _j^{\prime }\right)
\end{eqnarray}
respectively. This subroutine is used for all distributions except
$\alpha\alpha\alpha\alpha$, $\alpha\beta\alpha\beta$,
and $\beta\alpha\beta\alpha$.
In the latter cases, instead of subroutine
{\bf SSA} the subroutine {\bf SS1} is used.

\subsubsection{Spin--other--orbit interaction}

The subroutine {\bf SOOC} investigates the spin--other--orbit operator 
(\ref{eq:soo-a}).
%which according to (16) of \cite{method3} has the tensorial form
%\begin{eqnarray}
%\label{eq:m-b}
%   \D
%  H_{12}^{soo} \equiv \sum_k \left\{
%   H_{soo}^{\left( k-1k1,101\right) } + H_{soo}^{\left( k-1k1,011\right)
%  }+ H_{soo}^{\left( kk1,101\right) } + 
%   H_{soo}^{\left( kk1,011\right) }+ H_{soo}^{\left( k+1k1,101\right)
%   } + H_{soo}^{\left( k+1k1,011\right) }\right\}. \nonumber
%\end{eqnarray}
This subroutine is used for all distributions except
$\alpha\alpha\alpha\alpha$, $\alpha\beta\alpha\beta$,
$\beta\alpha\beta\alpha$, $\alpha\beta\beta\alpha$,
$\beta\alpha\alpha\beta$ and $\alpha\alpha\beta\beta$.
For distributions
$\alpha\alpha\alpha\alpha$, $\alpha\beta\alpha\beta$ and
$\beta\alpha\beta\alpha$,
according to (34), (39) and (40) of \cite{method3}, the tensorial form
\begin{eqnarray}
\label{eq:m-bb}
   \D
   H_{12}^{soo} \equiv \sum_k \left\{ H_{soo}^{\left( k-1k1,101\right) } + 
   H_{soo}^{\left( k-1k1,011\right)} + H_{soo}^{\left( k+1k1,101\right)} + 
   H_{soo}^{\left( k+1k1,011\right)}\right\} \nonumber
\end{eqnarray}
is valid. Therefore instead of subroutine
{\bf SSC} the subroutines {\bf SOO1111}, {\bf SOO1212}, {\bf SOO1212}
respectively are used. The distribution
$\alpha\beta\beta\alpha$, $\beta\alpha\alpha\beta$ and
$\alpha\alpha\beta\beta$
are also calculated according to separate expressions. These are treated by the
subroutines
{\bf SOO1221P}, {\bf SOO1221P} and {\bf SOO1122P}.

\medskip

The subroutines {\bf SOOA} and {\bf SOOB} investigate the
submatrix elements of the spin--other--orbit interaction operator
$$\left( n_i\lambda _in_j\lambda _j||H_{ss}^{\left( k-1 k 1,101 \right)
}||n_i^{\prime }\lambda _i^{\prime }n_j^{\prime }\lambda _j^{\prime }\right),
~~~~~
\left( n_i\lambda _in_j\lambda _j||H_{ss}^{\left( k-1 k 1,011 \right)
}||n_i^{\prime }\lambda _i^{\prime }n_j^{\prime }\lambda _j^{\prime }\right),$$

$$\left( n_i\lambda _in_j\lambda _j||H_{ss}^{\left( k k 1,101 \right)
}||n_i^{\prime }\lambda _i^{\prime }n_j^{\prime }\lambda _j^{\prime }\right),
~~~~~
\left( n_i\lambda _in_j\lambda _j||H_{ss}^{\left( k k 1,011 \right)
}||n_i^{\prime }\lambda _i^{\prime }n_j^{\prime }\lambda _j^{\prime }\right),$$

$$\left( n_i\lambda _in_j\lambda _j||H_{ss}^{\left( k+1 k 1,101 \right)
}||n_i^{\prime }\lambda _i^{\prime }n_j^{\prime }\lambda _j^{\prime }\right),
~~~~~
\left( n_i\lambda _in_j\lambda _j||H_{ss}^{\left( k+1 k 1,011 \right)
}||n_i^{\prime }\lambda _i^{\prime }n_j^{\prime }\lambda _j^{\prime }\right)$$
according to the formulas (26), (27) and (28) of \cite{method3} respectively.
These subroutines are used for all distributions except
$\alpha\alpha\alpha\alpha$. In the latter case instead of subroutines
{\bf SOOA} and {\bf SOOB} the subroutine {\bf SOO1} is used.

\section{The Library usage for other programs}

There are different versions of \texttt{ATSP\_MCHF} \cite{Fbook,F,new}, which support
non--relativistic atomic structure calculations. Each of them has some 
specific disparity. But it is possible to implement
the library \texttt{SAI} in all of them and in other
programs as well. The one example of such connection between \texttt{ATSP\_MCHF}
and \texttt{SAI} 
is shown for the program \texttt{Breit} from \cite{F} in Table~2. The modify 
package \texttt{ATSP\_MCHF}~\cite{F} which is fit to the 
module \texttt{SAI} is coming with the distribution of the library \texttt{SAI}.

\medskip

For writing the similar interface for other programmes like \cite{EJN}
we need to fill in the Common block \texttt{MEDEFN} from
\texttt{ATSP\_MCHF}, to write similar program as is presented in the Table~2 and 
to make some other subroutines similar as is presented in this
distribution (the files \texttt{breit.f}, \texttt{savenon.f} and \texttt{setupgg.f} 
in the directory \texttt{/dudu/breit}).

\medskip

\section*{Table 2. THE MODIFY PROGRAM \texttt{BREIT} FOR CONNECTION BETWEEN
\texttt{ATSP\_MCHF} AND \texttt{SAI}.}

%\begin{small}
\begin{scriptsize}
\begin{verbatim}
      PROGRAM BREITF
      IMPLICIT DOUBLE PRECISION(A-H,O-Z)
      PARAMETER (NWD=60,NCD=1000,NCD2=2*NCD)
      CHARACTER ANS*2, NAME(2)*24
      COMMON/DEBUG/IBUG1,IBUG2,IBUG3,NBUG6,NBUG7,IFULL
      COMMON/CONSTS/ZERO,TENTH,HALF,ONE,TWO,THREE,FOUR,SEVEN,ELEVEN,EPS
      COMMON/DIAGNL/IDIAG,JA,JB
      COMMON /FOUT/NOV(2),IOVLAP(10,2),NF,NG,NR,NL,NZ,NN,NV,NS,IFLAG,NIJ
      COMMON/IMAGNT/CONST,CONSOO,CONSS,ISPORB,ISOORB,ISPSPN,
     :     IREL,ISTRICT,IZOUT,IELST,ITENPR
      COMMON/INFORM/IREAD,IWRITE,IOUT,ISC(8)
      COMMON/MEDEFN/IHSH,NJ(16),LJ(16),NOSH1(16),NOSH2(16),J1QN1(31,3),
     :     J1QN2(31,3),IJFUL(16)
      COMMON/PHASES/SIGNFA(NCD2),ICSTAS
      COMMON/STATES/NCFG,NOCCSH(NCD2),NOCORB( 5,NCD2),NELCSH( 5,NCD2),
     :     J1QNRD( 9,NCD2),MAXORB,NJCOMP(NWD),LJCOMP(NWD),IAJCMP(NWD)
      COMMON /OPERAT/ ICOLOM,ISOTOP,IORBORB
      COMMON /BREIT/ ISPORBG,ISOORBG,ISPSPNG
      COMMON/STEGG/IX,IGGG,IRHO,ISIG,IRHOP,ISIGP
      DIMENSION NFLG(20),IRFST(NCD2),NCOUNT(8)
      EQUIVALENCE (NCOUNT(1),NF)
      LOGICAL INCL
  105 FORMAT (49H ISPORB=0 AND ISOORB=1 CAUSES THE PROGRAM TO FAIL,
     :  34H BECAUSE THE BLUME WATSON FORMULAE,/
     :  50H CANNOT BE USED FOR CLOSED SUBSHELLS.  TO OVERCOME,
     :  34H THIS, THE CODE HAS SET ISPORB = 1//)
   11 FORMAT(////' THE TYPE OF CALCULATION IS DEFINED BY ',
     :  'THE FOLLOWING PARAMETERS - '/
     : 5X,22H BREIT-PAULI OPERATORS,13X,8HIREL   =,I2/
     : 5X,27H PHASE CONVENTION PARAMETER,8X,8HICSTAS =,I2/)
   13 FORMAT(40H RELATIVISTIC OPERATORS TO BE INCLUDED -/5X,13H SPIN-ORB
     :IT (,I1,22H),  SPIN-OTHER-ORBIT (,I1,15H),  SPIN-SPIN (,I1,1H)/)
   24 FORMAT(36H0INITIAL DEBUG: IN 1-ELECTRON PART =,I2,2H ,,5X,
     : 20H IN 2-ELECTRON PART=,I2/16X,23HIN RECOUPLING PACKAGE =,I2/)
   42 FORMAT(//' MATRIX ELEMENTS CONSTRUCTED USING ',
     :       'THE SPHERICAL HARMONIC PHASE CONVENTION OF'/)
   43 FORMAT(16X,47HCONDON AND SHORTLEY, THEORY OF ATOMIC STRUCTURE/16X,
     :47H-----------------------------------------------///)
   44 FORMAT(25X,42HFANO AND RACAH, IRREDUCIBLE TENSORIAL SETS/25X,42H--
     :----------------------------------------///)
   50 FORMAT(/20X,'================================='/
     :        20X,' B R E I T - P A U L I      2001'/
     :        20X,'================================='/)
   78 FORMAT(19H DEBUG PARAMETERS -/5X,16H NBUG6(TENSOR) =,I2/5X,
     : ' NBUG7(RELATIVISTIC OPERATORS - SO,SOO,SS) =',I2//)
*
*     SET INPUT AND OUTPUT CHANNELS
      ISOTOP=0
      ICOLOM=0
      IORBORB=0
      IREAD=4
      IWRITE=6
      IOUT = 7
      DO 2 I = 1,8
         ISC(I) = 10 + I
         NCOUNT(I) = 0
         OPEN(UNIT=ISC(I),STATUS='SCRATCH',FORM='UNFORMATTED')
    2 CONTINUE
      NAME(1) = 'cfg.inp'
      NAME(2) = 'int.lst'
      OPEN(UNIT=IOUT,FILE=NAME(2),STATUS='UNKNOWN')
      NIJ = 0
      NHDEL=10
      MXIHSH=(10)
   10 WRITE(IWRITE,50)
      WRITE(0,'(A/A/A/A)') ' Indicate the type of calculation ',
     : ' 0 => non-relativistic Hamiltonian only;',
     : ' 1 => one or more relativistic operators only;',
     : ' 2 => non-relativistic operators and selected relativistic:  '
      READ(5,*) IREL
      IF (IREL.NE.1) ICOLOM=1
      WRITE(0,'(A)') ' Is full print-out requested? (Y/N) '
      READ(5,'(A2)') ANS
      IFULL = 0
      IF (ANS .EQ. 'Y' .OR. ANS .EQ. 'y') IFULL = 1
      WRITE(0,'(/A)')
     :    ' Phases:- Condon and Shortley or Fano and Racah ? (CS/FR) '
      READ(5,'(A2)') ANS
      IF ( ANS .EQ. 'CS' .OR. ANS .EQ. 'cs') THEN
         ICSTAS = 1
      ELSE
         ICSTAS = 0
      END IF
      ISPORB = 0
      ISOORB = 0
      ISPSPN = 0
      IORBORB=0
      IF (IREL .NE. 0) THEN
         ISPORB = 1
         ISOORB = 1
         ISPSPN = 1
         WRITE(0,'(A)') ' All relativistic operators ? (Y/N) '
         READ(5,'(A2)') ANS
         IF ( ANS .EQ. 'N ' .OR. ANS .EQ. 'n ') THEN
            WRITE(0,'(A)') ' Spin-orbit ? (Y/N) '
            READ(5,'(A2)') ANS
            IF ( ANS .EQ. 'N ' .OR. ANS .EQ. 'n ') ISPORB = 0
            WRITE(0,'(A)') ' Spin-other-orbit ? (Y/N) '
            READ(5,'(A2)') ANS
            IF ( ANS .EQ. 'N ' .OR. ANS .EQ. 'n ') ISOORB = 0
            WRITE(0,'(A)') ' Spin-spin ? (Y/N) '
            READ(5,'(A2)') ANS
            IF ( ANS .EQ. 'N ' .OR. ANS .EQ. 'n ') ISPSPN = 0
         END IF
         IF(ISPORB.EQ.0.AND.ISOORB.NE.0) THEN
            WRITE (IWRITE,105)
         END IF
      END IF
      IBUG1 = 0
      IBUG2 = 0
      IBUG3 = 0
      NBUG6 = 0
      NBUG7 = 0
      WRITE(IWRITE,11) IREL,ICSTAS
      ISPORBG = ISPORB  
      ISOORBG = ISOORB 
      ISPSPNG = ISPSPN 
      CALL FACTRL(32)
*
* --- READ IN THE SET OF CONFIGURATIONS
      CALL ACNFIG(NAME(1))
      IDG = 0
      NZERO = NCFG
      NEW = NCFG
      DO 554 I = 1,NZERO
554   IRFST(I) = 1
      ISTART = 1
      WRITE(0,'(A)') ' All Interactions? (Y/N): '
      READ (5,'(A2)') ANS
      IF (ANS .NE. 'Y' .AND. ANS .NE. 'y') THEN
          WRITE(0,'(A,I3,A/A)') ' Of the ',NCFG,
     :         ' configurations, how many at the end are new? ',
     :         ' How many configurations define the zero-order set?'
         READ (5,*) NEW,NZERO
	 IF (NEW .EQ. 0) NEW = NCFG
         ISTART = NCFG - NEW + 1
         IF (NZERO .eq. 0) NZERO = NCFG
         IF (IREL .NE. 0 .AND. NZERO .NE. NCFG) THEN
            WRITE(0,'(A)') ' Rel interaction with all the zero block? '
            READ (5,'(A2)') ANS
            IF (ANS .NE. 'Y' .AND. ANS .NE. 'y') THEN
               DO 553 I = 1,(20)
553            NFLG(I) = 0
               WRITE(0,*) ' Define your reference set : FORM(20I3)'
               READ (5,'(20I3)') (NFLG(I),I=1,(20))
               DO 551 I = 1,(ISTART-1)
               IRFST(I) = 0
               DO 552 J = 1,(20)
               IF (NFLG(J) .EQ. I) THEN
                  IRFST(I) = 1
                  GO TO 552
               ENDIF
552            CONTINUE
551            CONTINUE
            ENDIF
            WRITE(0,*) ' Diagonal rel corrections ? (Y/N): '
            READ (5,'(A2)') ANS
            IF (ANS.EQ.'Y'.OR.ANS.EQ.'y') IDG = 1
         ENDIF
         ISTRICT = 1
         WRITE(0,*)  ' Restricted Two-body interactions? (Y/N); '
         READ (5,'(A2)') ANS
         IF (ANS .NE. 'Y' .and. ANS .NE. 'y' ) ISTRICT = 0
      END IF
*
* ...  Start the calculation
      DO 6 JA = ISTART, NCFG
      ICOUNT = 0
*
* ... Set up defining quantum numbers for each matrix element.
      CALL SETUP(JA,JB)
      IF(IBUG1.NE.0 .OR. IBUG2.NE.0) CALL VIJOUT(JA,JB)
      IF(IHSH.GT.MXIHSH) STOP
*
* ... TEST ON POSSIBLE RECOUPLING ORTHOGONALITY.
      CALL ORTHOG(LET,INCL)
      IF (LET .NE. 0) THEN
        CALL SETUPGG
        IF (IX .LE. 4) THEN
          IF (IFULL .NE. 0) WRITE(IWRITE,77)
   77     FORMAT(///30X,'MULTIPLYING FACTOR',11X,'TYPE OF INTEGRAL')
          CALL NONBP
          IF (IFLAG .NE. 0) NIJ = NIJ + 1
        ENDIF
      ENDIF
    7 CONTINUE
    6 CONTINUE
      NTOTAL = ((2*NCFG - NEW +1)*NEW)/2
      WRITE(0,*) NTOTAL, ' matrix elements'
      WRITE(0,*) NIJ, ' non-zero matrix elements'
      WRITE(0,*) 100*NIJ/REAL(NTOTAL),' % dense'
      WRITE(0,*) 'NF=',nf,' NG=',ng,' NR=',nr,' NL=',nl
      WRITE(0,*) 'NZ=',nz,' NN=',nn,' NV=',nv,' NS=',ns
      WRITE(0,*) 'Total number of terms =', NF+NG+NR+NL+NZ+NN+NV+NS
      CALL OUTLSJ
      WRITE(IWRITE,42)
      IF(ICSTAS.EQ.0) THEN
         WRITE(IWRITE,44)
        ELSE
         WRITE(IWRITE,43)
      END IF
      STOP
      END
\end{verbatim}
%\end{small}
\end{scriptsize}

\section{Installation of the Library}

The library \texttt{SAI} will be distributed as compressed \texttt{tar} archive file.
On a UNIX (or compatible) workstation, the commands \texttt{uncompress targ\_lib\_SAI} 
and \texttt{tar xvf targ\_lib\_SAI} reconstruct the file structure. The directory
\texttt{LIBRARY\_SAI} contains the source code
(subdirectories \texttt{dudu}, \texttt{nore}, \texttt{recls}, \texttt{sqlsf1}
and \texttt{sqlsf1}) as well as makefile \texttt{make\_lib\_SAI}.
It also includes a modify 
package \texttt{ATSP\_MCHF}~\cite{F} which is fited to the 
module \texttt{SAI}.

\medskip

The command \texttt{make\_lib\_SAI} creates the libraries \texttt{libdudu.a},
\texttt{libnore.a}, \texttt{librecls.a}, \texttt{libsqlsf1.a} and
\texttt{libsqlsf2.a} in directory \texttt{bin}.

\medskip

\section{Examples}

The way to determine the ionization potential for $Ce$ is demonstrated in this section. 
For this, one needs to calculate the ground state energy of $Ce$ atom ($^{1}G$), 
the ground state of a $Ce^{+}$ ion ($^{2}G$), and find the difference.
It has been shown in ~\cite{Tatewaki,GFGR:99,ER_ion}
that the non--relativistic
Hartree--Fock ionization potential of singly ionized lanthanides follows the
observed trends.
Therefore we have chosen the non--relativistic approximation, and find
the energy values by solving the multi--configuration Hartree--Fock equations.
For brevity, we present only the $Ce^{+}$ test cases. The energy for
$Ce$ is found in the same way.

\medskip

In Table~3 we present an example of generating a set of configurations
where the ground state energy of th $Ce^{+}$ ion is to be calculated by the 
multiconfiguration Hartree--Fock method.
The configurations are obtained by taking the singles and doubles excitations 
from $\left \{ 4f, 5d, 6s \right\}$ to the active set 
$\left \{ 5f, 5g \right\}$ (since neither $5d$ nor $6s$ are filled shells, 
better accuracy would be obtained if these orbitals
were also included in the active set)~\cite{ER_ion}.
In this case the program \texttt{GENCL}~\cite{FL} generates a $cfg.out$ file 
with 40 configuration states.

\section*{Table 3. TEST RUN OUTPUT}

\begin{scriptsize}
%\small
\begin{verbatim}

>>genclf
       -----------------------------------------------------------
       You are under the program GENCL
              which GENerates a Configuration List
       Type H (Help) to obtain information about the input format
       Type <RETURN> if you already know
       --------------------------------------------------------
              Header  ?
> Ce+ ground state
       Closed Shells  ?
> 1s  2s  2p  3s  3p  3d  4s  4p  4d  5s  5p
       Reference Set  ?
> 4f(1)5d(1)6s(1)
                 2  ?
>
          Active Set  ?
>
        Replacements  ?
> sd
         Virtual Set  ?
> 5f,5g
    From which shell  ?
> 1
      To which shell  ?
> 3
         Final Terms  ?
> 2G
                 2  ?
>
    ***************          I N P U T  D A T A          **********
              Header  :  Ce+ ground sate
        Closed shell  :   1s  2s  2p  3s  3p  3d  4s  4p  4d  5s  5p
       Reference Set  :  4f(1)5d(1)6s(1)
          Active Set  :
        Replacements  :  sd
         Virtual Set  :  5f,5g
              From which shell  :  1
                To which shell  :  3
         Final Terms  :  2G
  GENERATE ALL COUPLINGS FOR EACH MEMBER OF THE REFERENCE SET
         -------------       Ce+ ground state   --------
          Closed Shells  :   1s  2s  2p  3s  3p  3d  4s  4p  4d  5s  5p 
          Configuration  :  4f( 1)  5d( 1)  6s( 1)
        Their couplings  :       2F1     2D1     2S1     1G0     2G0
                                 2F1     2D1     2S1     3G0     2G0
     FOR VIRTUAL SET GENERATE CONFIGURATION STATES FOR S-REPLACEMENT
         -------------       Ce+ ground state   --------
          Closed Shells  :   1s  2s  2p  3s  3p  3d  4s  4p  4d  5s  5p 
          Reference set  :  4f(1)5d(1)6s(1)                                             
            Virtual Set  :  5f,5g                                                       
          S-Replacement  :  4f  = 5f                                                    
          Configuration  :  5d( 1)  6s( 1)  5f( 1)
        Their couplings  :       2D1     2S1     2F1     1D0     2G0
                                 2D1     2S1     2F1     3D0     2G0
... Output omitted for brevity ...
    FOR VIRTUAL SET, GENERATE CONFIGURATION STATES FOR D-REPLACEMENT
         -------------       Ce+ ground state   --------
          Closed Shells  :   1s  2s  2p  3s  3p  3d  4s  4p  4d  5s  5p 
          Reference set  :  4f(1)5d(1)6s(1)                                             
            Virtual Set  :  5f,5g                                                       
          S-Replacement  :  4f .5d  =5f .5g                                             
          Configuration  :  6s( 1)  5f( 1)  5g( 1)
        Their couplings  :       2S1     2F1     2G1     1F0     2G0
                                 2S1     2F1     2G1     3F0     2G0
... Output omitted for brevity ...
          OK!
          List of configurations and their couplings
          is in the file   cfg.inp
\end{verbatim}
\end{scriptsize}

\medskip

The generating of the angles file $int.lst$ is demonstrated in Table 4 with the
modify package \texttt{ATSP\_MCHF}~\cite{F} which is fited to the 
module \texttt{SAI} and presented in the
distribution.
It will be needed for calculating the energy of $^{2}G$ state of
$Ce^{+}$ with the \texttt{MCHF} program~\cite{MSHF:91}. The program input file $cfg.inp$
is taken with the help of renaming $cfg.out$ to $cfg.inp$.

\medskip

\section*{Table 4. TEST RUN OUTPUT}

\begin{scriptsize}
\begin{verbatim}
>>breitf                                   ========> Obtain Expressions for
                                                     Breit-Pauli Interactions

                    =======================
                     B R E I T - P A U L I
                    =======================

 Indicate the type of calculation
 0 => non-relativistic Hamiltonian only;
 1 => one or more relativistic operators only;
 2 => non-relativistic operators and selected relativistic:
>0
 Is full print-out requested? (Y/N)
>N
 Phases:- Condon and Shortley of Fano and Racah ? (CS/RF)
>CS
 All Interactions? (Y/N)
>Y

 THE TYPE OF CALCULATION IS DEFINED BY THE FOLLOWING PARAMETERS - 
      BREIT-PAULI OPERATORS             IREL   = 0
      PHASE CONVENTION PARAMETER        ICSTAS = 1

                             ---------------------
                             THE CONFIGURATION SET
                             ---------------------

 STATE  (WITH 40 CONFIGURATIONS):
 -------------------------------

 THERE ARE  5 ORBITALS AS FOLLOWS:

       4f  5d  6s  5f  5g

 THERE ARE 11 CLOSED SUBSHELLS COMMON TO ALL CONFIGURATIONS AS FOLLOWS:

       1s  2s  2p  3s  3p  3d  4s  4p  4d  5s  5p

 CONFIGURATION  1 ( OCCUPIED ORBITALS= 3 ):  4f( 1)  5d( 1)  6s( 1)
                           COUPLING SCHEME:     2F1     2D1     2S1
                                                           1G0     2G0
... Output omitted for brevity ...

 CONFIGURATION 40 ( OCCUPIED ORBITALS= 2 ):  4f( 1)  5g( 2)
                           COUPLING SCHEME:     2F1     3H2
                                                           2G0

 MATRIX ELEMENTS CONSTRUCTED USING THE SPHERICAL HARMONIC PHASE CONVENTION OF
                CONDON AND SHORTLEY, THEORY OF ATOMIC STRUCTURE
                -----------------------------------------------
>>cat int.lst                          ===> Display the int.lst file produced
Ce+ ground sate
                                                                        
 F 0( 4f, 5d)    12              R 0( 4f 5d, 5f 5d)    14          R 5( 5d 5g, 5f 5f)  2008
 F 0( 4f, 6s)    16              R 0( 4f 6s, 5f 6s)    20          R 5( 5f 5f, 5g 5g)  2014
 F 0( 4f, 5f)    22              R 0( 4f 5g, 5f 5g)    32          R 6( 4f 5d, 5f 5g)  2018
 F 0( 4f, 5g)    40              R 1( 4f 5d, 5d 5f)    46          R 6( 4f 5g, 4f 5d)  2082
 F 0( 5d, 6s)    44              R 1( 4f 5d, 5g 5f)    50          R 6( 4f 5g, 5f 5d)  2146
 F 0( 5d, 5f)    56              R 1( 4f 5f, 5d 5g)   110          R 6( 4f 5g, 5f 5g)  2178
 F 0( 5d, 5g)    76              R 1( 4f 5f, 5g 5d)   124          R 6( 5d 5f, 5g 5f)  2182
 F 0( 6s, 5f)    80              R 1( 4f 5g, 5d 4f)   188          R 6( 5d 5g, 5g 5g)  2232
 F 0( 6s, 5g)    84              R 1( 4f 5g, 5d 5f)   250          R 7( 4f 5g, 5g 5f)  2336
 F 0( 5f, 5f)    90              R 1( 4f 5g, 5g 5f)   354          R 7( 5f 5f, 5g 5g)  2342
 F 0( 5f, 5g)   102              R 1( 5d 5f, 5f 5g)   356          *
 F 0( 5g, 5g)   108              R 1( 5d 5g, 5f 5f)   406            -1.00000000R 26 14
 F 2( 4f, 5d)   120              R 1( 5f 5f, 5g 5g)   412             1.00000000R 27 15
 F 2( 4f, 5f)   130              R 2( 4f 5d, 4f 6s)   414             1.00000000R 23 11
 F 2( 4f, 5g)   166              R 2( 4f 5d, 5f 5d)   428             0.50000000R  3  1
 F 2( 5d, 5f)   178              R 2( 4f 5d, 5f 6s)   430             0.86602540R  4  1
 F 2( 5d, 5g)   226              R 2( 4f 5d, 5f 5g)   434            -1.00000000R 22 10
 F 2( 5f, 5f)   232              R 2( 4f 6s, 5f 5d)   436             1.00000000R 21  9
 F 2( 5f, 5g)   258              R 2( 4f 5g, 4f 5d)   500             1.00000000R 19  7
 F 2( 5g, 5g)   264              R 2( 4f 5g, 5f 5d)   564            -1.00000000R 20  8
 F 4( 4f, 5d)   276              R 2( 4f 5g, 5f 5g)   604            -1.00000000R 28 16
 F 4( 4f, 5f)   287              R 2( 5d 5f, 6s 5f)   606             1.00000000R 25 13
 F 4( 4f, 5g)   324              R 2( 5d 5f, 5g 5f)   610            -0.86602540R  3  2
 F 4( 5d, 5f)   336              R 2( 5d 5g, 6s 5d)   650             0.50000000R  4  2
 F 4( 5d, 5g)   388              R 2( 5d 5g, 6s 5g)   670            -1.00000000R 24 12
 F 4( 5f, 5f)   394              R 2( 5d 5g, 5g 5g)   720            -0.86602540R  3  2
 F 4( 5f, 5g)   420              R 3( 4f 5d, 5d 5f)   734             0.50000000R  3  1
 F 4( 5g, 5g)   426              R 3( 4f 5d, 6s 4f)   736             0.86602540R  4  1
 F 6( 4f, 5f)   435              R 3( 4f 5d, 6s 5f)   738            -1.00000000R 18  6
 F 6( 4f, 5g)   464              R 3( 4f 5d, 5g 5f)   742             0.50000000R  4  2
 F 6( 5f, 5f)   470              R 3( 4f 6s, 5d 5f)   744             1.00000000R 17  5
 F 6( 5f, 5g)   490              R 3( 4f 6s, 6s 5f)   748            -1.00000000R 18  6
 F 6( 5g, 5g)   496              R 3( 4f 6s, 5g 5f)   768             1.00000000R 25 13
 F 8( 5g, 5g)   502              R 3( 4f 5f, 5d 6s)   780             1.00000000R 17  5
 *                               R 3( 4f 5f, 5d 5g)   836             1.00000000R 19  7
    1.00000000F  1  1            R 3( 4f 5f, 6s 5d)   848            -1.00000000R 20  8
    1.00000000F 13 13            R 3( 4f 5f, 6s 5g)   860            -1.00000000R 24 12
    1.00000000F 14 14            R 3( 4f 5f, 5g 5d)   884             1.00000000R 21  9
    1.00000000F  7  7            R 3( 4f 5f, 5g 6s)   890            -1.00000000R 26 14
    1.00000000F 15 15            R 3( 4f 5g, 5d 4f)   942            -1.00000000R 28 16
(cont.)                          R 3( 4f 5g, 5d 5f)   992            -1.00000000R 22 10
              *                  R 3( 4f 5g, 6s 4f)  1024             1.00000000R 23 11
 G 0( 4f, 5f)    21              R 3( 4f 5g, 6s 5f)  1046             1.00000000R 27 15
 G 1( 4f, 5d)    33              R 3( 4f 5g, 5g 5f)  1150             0.17142857R 21  9
 G 1( 4f, 5g)   111              R 3( 5d 6s, 5f 5f)  1152             0.02857143R 19  7
 G 1( 5d, 5f)   124              R 3( 5d 5f, 5f 6s)  1154             0.14285714R  4  2
 G 1( 5f, 5g)   182              R 3( 5d 5f, 5f 5g)  1156            -0.14285714R  3  1
 G 2( 4f, 5f)   202              R 3( 5d 5g, 5f 5f)  1206             0.08571429R 20  8
 G 2( 5d, 6s)   207              R 3( 6s 5f, 5f 5g)  1216             0.28571429R 22 10
 G 2( 5d, 5g)   317              R 3( 6s 5g, 5f 5f)  1218             0.42857143R 28 16
 G 3( 4f, 5d)   329              R 3( 5f 5f, 5g 5g)  1224             0.28571429R 27 15
 G 3( 4f, 6s)   334              R 4( 4f 5d, 5f 5d)  1238            -0.24743583R  4  1
 G 3( 4f, 5g)   413              R 4( 4f 5d, 5f 5g)  1242             0.02857143R 24 12
 G 3( 5d, 5f)   426              R 4( 4f 6s, 5f 5g)  1252             0.42857143R 23 11
 G 3( 6s, 5f)   431              R 4( 4f 5g, 4f 5d)  1316             0.17142857R 26 14
 G 3( 5f, 5g)   489              R 4( 4f 5g, 4f 6s)  1338            -0.24743583R  3  2
 G 4( 4f, 5f)   510              R 4( 4f 5g, 5f 5d)  1402             0.08571429R 25 13
 G 4( 5d, 5g)   606              R 4( 4f 5g, 5f 6s)  1434            -0.14982984R 17  1
 G 4( 6s, 5g)   612              R 4( 4f 5g, 5f 5g)  1476             0.25951289R 17  2
 G 5( 4f, 5d)   624              R 4( 5d 6s, 5g 5g)  1478             0.14982984R 18  2
 G 5( 4f, 5g)   703              R 4( 5d 5f, 5g 5f)  1482             0.25951289R 18  1
 G 5( 5d, 5f)   716              R 4( 5d 5g, 5d 6s)  1512            -0.03937627R 30 28
 G 5( 5f, 5g)   774              R 4( 5d 5g, 5g 6s)  1552            -0.08460778R 32 25
 G 6( 4f, 5f)   795              R 4( 5d 5g, 5g 5g)  1602             0.11533906R 33 24
 G 6( 5d, 5g)   905              R 4( 6s 5f, 5g 5f)  1622             0.00067076R 34 28
 G 7( 4f, 5g)   984              R 4( 6s 5g, 5g 5g)  1624            -0.05155563R 30 27
 G 7( 5f, 5g)  1042              R 5( 4f 5d, 5d 5f)  1638            -0.04956348R 33 23
 *                               R 5( 4f 5d, 5g 5f)  1642             0.12698464R 30 26
   -0.27380952G 34 34            R 5( 4f 5f, 5d 5g)  1702             0.09770956R 30 25
   -0.50787450G 34 33            R 5( 4f 5f, 5g 5d)  1726             0.12978765R 33 22
    0.55135887G 34 32            R 5( 4f 5g, 5d 4f)  1790            -0.21161602R 30 24
   -0.09731237G 34 31            R 5( 4f 5g, 5d 5f)  1852            -0.12989160R 33 25
   -1.00946357G 34 30            R 5( 4f 5g, 5g 5f)  1956             0.02273390R 30 23
(cont.)                          R 5( 5d 5f, 5f 5g)  1958          (cont.)
\end{verbatim}
\end{scriptsize}

\medskip

The energy of $^{2}G$ state of $Ce^{+}$ is found by the
multi--configurational Hartree--Fock method. First, the wave functions are
obtained by the HF method, using the \texttt{HF96} program
(see Gaigalas and Fischer~\cite{GFa}).
These functions are used as starting ones in solving the multi--configurational
Hartree--Fock equations. The input files for \texttt{MCHF} program~\cite{MSHF:91}
$int.lst$, and $cfg.inp$, are taken from the examples described above.

\section{Conclusion}

The library \texttt{SAI} supports large scale 
computations of
open--shell atoms using multi--configuration Hartree--Fock or configuration interaction
approaches and may be useful for developing codes for calculating the spin--angular parts of
effective operators from many--body perturbation theory and orthogonal operators
or for evaluating the relativistic
Hamiltonian in $LS$--coupling as well as for various versions of semi--empirical methods.
For example, it
expands the use of possibilities of package 
\texttt{MCHF\_ASP}~\cite{Fbook,F,new} 
because the new program is faster and provides coefficients and matrix
elements for shells ($nl$) with $l=0$, $1$, $2$ and $3$, 
and $l^{2}$ for $l \ge 3$. It is important to underline that the use of the library described
allows to perform calculations of energy spectra of practical any atom or ion starting with 
$LS$--coupling and accounting for all relativistic corrections of the order $\alpha ^2$.

\medskip

Program is obtainable from the Institute of Theoretical Physics and Astronomy,
A. Go\v stauto 12, Vilnius, 2600, LITHUANIA.~~ E--mail: gaigalas@itpa.lt.

\section*{Acknowledgments}

The author is grateful to Prof. Z. Rudzikas for encouraging and valuable
suggestions.

\clearpage
\end{document}